# Crochets de Poisson, théories de jauge et quantification

Winston FAIRBAIRN et Catherine MEUSBURGER

Nous présentons une étude des systèmes avec contraintes et de leur quantification, montrant l'importance du rôle des crochets de Poisson dans les théories de jauge.

## Théorie classique et quantification

Depuis leur invention il y a deux siècles, les crochets de Poisson n'ont cessé de jouer un rôle fondamental dans divers domaines de la physique mathématique. Ils demeurent encore aujourd'hui un outil indispensable de la recherche en physique mathématique. Nous chercherons à illustrer l'importance des travaux de Poisson dans un domaine particulier, celui des interactions fondamentales. Un des problèmes centraux dans ce domaine est l'analyse du comportement des interactions à l'échelle microscopique, où l'approximation de la mécanique classique cesse d'être une description valable de la nature. En d'autres termes, il s'agit de quantifier la théorie classique, c'est-à-dire d'obtenir une description de ces interactions dans le cadre de la mécanique quantique.

Pour trois des quatre interactions fondamentales, l'electromagnétisme, et les interactions nucléaires faible et forte, il existe une description quantique, le modèle standard de la physique des particules, qui constitue un modèle unifié, vérifié expérimentalement avec une précision remarquable. Il reste cependant un certain nombre de problèmes ouverts qui sont à l'origine d'efforts de recherche importants en physique mathématique contemporaine. Citons par exemple le problème du confinement, le problème d'existence d'une théorie quantique des champs fondée sur la théorie de Yang et Mills et celui d'un gap de masse, ainsi que la classification des structures différentielles des variétés de dimension quatre.

La quatrième et dernière interaction fondamentale, la gravitation, ne possède pas de description quantique pleinement cohérente. Ceci implique en particulier que notre description de l'univers cesse d'être valable pour des domaines d'énergie supérieurs à un certain seuil, appelé énergie de Planck, à partir duquel les effets quantiques de la gravité deviennent non négligeables. La résolution du problème de la quantification de la gravitation est l'un des défis fondamentaux de la physique théorique moderne, et malgré des progrès importants obtenus dans le cadre de l'approche dite non-perturbative à la gravité quantique et de la théorie des cordes, ce problème reste à ce jour non résolu.

Une des difficultés essentielle apparaissant dans le problème de quantification des quatre interactions fondamentales est que celles-ci sont décrites classiquement dans une formulation lagrangienne où l'action, c'est-à-dire l'intégrale temporelle de la fonction lagrangienne qui décrit le système, est invariante par certaines transformations, appelées transformations de jauge. Ces systèmes sont déjà complexes dans leur description classique car l'évolution temporelle y est indéterminée. L'invariance par les transformations de jauge se traduit par la présence de contraintes, et par conséquent de tels systèmes sont appelés systèmes avec contraintes. Le crochet de Poisson joue un rôle essentiel dans l'étude des théories physiques correspondantes, appelées théories de jauge, car il permet de déterminer et de classifier toutes les contraintes au niveau classique.

La quantification des systèmes avec contraintes repose sur une extension non triviale des méthodes habituelles s'appliquant aux systèmes plus élémentaires ne possédant pas d'invariance de jauge. Le crochet de Poisson permet de dicter la procédure de quantification en déterminant, en particulier, comment construire une évolution temporelle quantique. Il existe deux approches à la quantification d'un système classique, basées respectivement sur l'analyse lagrangienne utilisant des méthodes d'intégrales fonctionnelles, et sur le formalisme hamiltonien. Afin de comprendre la quantification des théories de jauge, nous nous intéresserons à la seconde méthode, c'est-à-dire à la quantification hamiltonienne, dite quantification canonique, qui nécessite au préalable une étude fine des propriétés du système hamiltonien classique. Il existe plusieurs méthodes de quantification canonique ; nous décrirons brièvement la réduction de l'espace des phases, puis la procédure de quantification de Dirac.

## Systèmes hamiltoniens avec contraintes

### L'invariance de jauge

Dans une théorie de jauge, l'intégrale temporelle du lagrangien est invariante par un groupe de transformations, les transformations de jauge. Ces transformations, dites locales, se distinguent des symétries globales usuelles des systèmes physiques en ce qu'elles dépendent du point d'espace-temps considéré. La présence de transformations de jauge manifeste l'existence de degrés de liberté superflus dans le lagrangien : le système considéré est décrit par un nombre de variables supérieur au nombre de degrés de liberté du système. Les quantités physiques associées aux degrés de liberté physiques du système sont les quantités dites observables, et elles sont invariantes par les transformations de jauge. En d'autres termes, les transformations de jauge sont des transformations qui relient deux descriptions équivalentes d'un même état physique.

La première découverte d'une théorie possédant une invariance de jauge date de 1864 lorsque James Clerk Maxwell (1831–1879) formula la théorie de l'électromagnétisme[1]. Il ignorait bien sûr cette notion. Ce fut Hermann Weyl (1885-1955) qui introduisit en 1918 la notion d'invariance d'échelle[2], appelée ensuite invariance de jauge, et ce fut en 1929 qu'il publia l'article considéré comme le texte fondateur de la théorie[3]. En 1954, C.N. Yang et R.L. Mills[4] eurent l'idée de généraliser l'invariance de jauge de l'électromagnétisme afin de décrire l'interaction forte entre particules élémentaires. C'est à partir de leurs travaux que fut élaboré, au milieu des années 1970, le modèle standard de la physique des particules[5], théorie de jauge qui est fondée sur la théorie des groupes de Lie et qui décrit toutes les interactions fondamentales sauf la gravitation.

Le principe fondamental sous-jacent à la théorie de la relativité générale d'Einstein[6] de 1915 est également un principe d'invariance de jauge, la covariance générale. Cependant, à la différence du cas des autres interactions, le groupe des transformations de jauge est là un groupe de dimension infinie, le groupe des difféomorphismes de la variété d'espace-temps, qui agit sur l'espace-temps lui-même et pas seulement sur les variables de champ de la théorie.

L'invariance de jauge est aujourd'hui universellement reconnue comme une pro-priété essentielle de toute théorie physique décrivant les interactions fondamentales, et l'étude des systèmes invariants par transformations de jauge est par conséquent d'importance capitale.

---

[1] J.C. Maxwell, A dynamical theory of the electromagnetic field, *Philosophical Transactions of the Royal Society of London*, 155 (1865), p. 459-512.

[2] H. Weyl, Gravitation und Elektrizität, *Sitzungsberichte der Königlich-preussischen Akademie der Wissenschaften zu Berlin*, 26 (1918), p. 465-480 ; réimprimé dans ses *Gesammelte Abhandlungen*, vol. 2, p. 29-42 (avec un commnentaire par Einstein et un texte écrit par Weyl en 1955).

[3] Elektron und Gravitation, *Zeitschrift für Physik*, 56 (1929), p. 330-352 ; *Gesammelte Abhandlungen*, vol. 3, p. 245-267. Sur l'histoire des débuts de la théorie de jauge, voir Lochlainn O'Raifeartaigh, *The Dawning of Gauge Theory*, Princeton, Princeton University Press, 1997.

[4] C. N. Yang et R. Mills, Conservation of isotopic spin and isotopic gauge invariance, *Physical Review*, 96 (1954), p. 191-195.

[5] Voir, par exemple, S.L. Glashow, Partial symmetries of weak interactions, *Nuclear Physics*, 22 (1961) p. 579-588 ; S. Weinberg, A model of leptons, *Physical Review Letters*, 19 (1967), p. 1264-1266 ; A. Salam, Weak and electromagnetic interactions, dans *Elementary Particle Physics : Relativistic Groups and Analyticity*, Eighth Nobel Symposium, éd. N. Svartholm, Stockholm, Almqvist and Wiksell, 1968, p. 367-377.

[6] A. Einstein, Grundlage der allgemeinen Relativitätstheorie, *Annalen der Physik*, 49 (1916), p. 769-822.

## Symétries de jauge dans les systèmes hamiltoniens

Dans les années 1950, P.A.M. Dirac (1902–1984) était à la recherche d'une formulation de l'électrodynamique quantique, ce qui l'amena à étendre la théorie hamiltonienne aux systèmes avec contraintes[7]. James L. Anderson, Peter G. Bergmann et Joshua N. Goldberg[8] contribuèrent également au développement de ces généralisations. Afin de simplifier l'exposition, nous considérons ici un système avec un nombre fini de degrés de liberté.

**La transformation de Legendre.** L'espace de configuration d'un système physique est modélisé par une variété différentielle $\mathcal{C}$. On considère la réunion des espaces vectoriels tangents en chaque point de la variété, appelée fibré tangent à $\mathcal{C}$ et notée $T(\mathcal{C})$. Un point dans $T(\mathcal{C})$ s'écrit $(q, v)$, où $q$ représente la position et $v$ la vitesse. Le système est décrit par un lagrangien, qui est une fonction sur $T(\mathcal{C})$ à valeurs réelles, notée $L$. On associe au lagrangien l'action, $S_L = \int_{t_1}^{t_2} dt\, L(q, v)$, et la formulation lagrangienne de la dynamique associée est obtenue en considérant le problème variationnel correspondant, c'est-à-dire en cherchant à minimiser l'action.

D'autre part, le formalisme hamiltonien de la dynamique du système s'exprime dans l'espace des phases, qui est le fibré cotangent de l'espace de configuration, noté $\mathcal{P} = T^*(\mathcal{C})$. En chaque point de l'espace de configuration, l'espace cotangent est le dual de l'espace vectoriel tangent ; un point dans $\mathcal{P}$ s'écrira donc $(q, p)$, où $p$ désigne le moment conjugué à la vitesse. De par sa structure de fibré cotangent, $\mathcal{P}$ est naturellement muni d'une structure symplectique. En effet, il existe sur tout fibré cotangent une forme différentielle de degré 1, la forme de Liouville, définie globalement. La différentielle de la forme de Liouville est une forme $\omega$ de degré 2, fermée et non dégénérée, ce qui par définition est une forme symplectique et donne donc à l'espace des phases une structure symplectique. On considère la structure de Poisson associée. Localement, dans un système de coordonnées défini par $q^a$ et $p_a$, $a = 1, \ldots, d$, où $d$ est la dimension de la variété de configuration, le crochet de Poisson de tout couple

---

[7] P.A.M. Dirac, Generalized Hamiltonian dynamics, *Canadian Journal of Mathematics*, 2 (1950), p. 129-148 ; The Hamiltonian form of field dynamics, *ibid.*, 3 (1951), p. 1-23 ; Generalized Hamiltonian dynamics, *Proceedings of the Royal Society* (London), A246 (1958), p. 326-332.

[8] J.L. Anderson et P.G. Bergmann, Constraints in covariant field theories, *Physical Review*, 83 (1951), p. 1018-1025 ; P.G. Bergmann, and I. Goldberg, Dirac bracket transformation in phase space. *Physical Review*, 98 (1955), p. 531-538. Pour un exposé général, voir le livre de Marc Henneaux et Claudio Teitelboim, *Quantization of Gauge Systems*, Princeton, Princeton University Press, 1992.

de fonctions $f$ et $g$ sur $\mathcal{P}$ est :

$$\{f,g\} = \sum_{a=1}^{d} (\frac{\partial f}{\partial q^a}\frac{\partial g}{\partial p_a} - \frac{\partial g}{\partial q^a}\frac{\partial f}{\partial p_a}).$$

On dit que les fonctions sur l'espace des phases forment une algèbre de Poisson.

La transition entre formalismes lagrangien et hamiltonien s'effectue à travers la transformation de Legendre qui est une application du fibré tangent $T(\mathcal{C})$ dans l'espace des phases $T^*(\mathcal{C})$. Cette application dépend du choix d'un lagrangien, c'est-à-dire d'une dynamique. Localement, elle est définie par

$$p_a = \frac{\partial L}{\partial v^a}, \qquad a = 1, \ldots, d.$$

**Contraintes primaires.** Il se peut que la transformation de Legendre, en tant qu'application entre variétés différentielles, ne soit pas un difféomorphisme, en particulier qu'elle ne soit pas surjective. En raisonnant localement en termes de l'application linéaire tangente associée, on voit que la transformation de Legendre est un difféomorphisme si le déterminant des dérivées secondes du lagrangien par rapport aux vitesses est non nul. Si, au contraire, ce déterminant est nul, le lagrangien est dit singulier. Il est alors impossible d'exprimer les vitesses en fonction des positions et des moments. En d'autres termes, les moments ne sont pas tous indépendants, et il existe un certain nombre $I$ de relations,

$$\phi_i(q,p) = 0, \qquad i = 1, ..., I.$$

Ces relations, qui sont la conséquence des propriétés d'invariance du lagrangien, s'appellent des contraintes primaires. La qualification « primaire » provient du fait que les équations du mouvement n'ont pas été utilisées pour obtenir les contraintes. Lorsque ces $I$ relations sont indépendantes, on parle de contraintes irréductibles. Nous ferons ici l'hypothèse que les contraintes sont irréductibles.

On voit ainsi que, dans le cas d'un lagrangien singulier, l'image du fibré tangent par la transformation de Legendre n'est pas l'espace des phases entier, mais un sous-ensemble de l'espace des phases appelé sous-variété des contraintes ou surface des contraintes[9], défini par les conditions $\phi_i(q,p) = 0$, $i = 1, \ldots, I$, que nous noterons $\Sigma$. Ce défaut de surjectivité implique que l'inverse de la transformation de Legendre est multivaluée. Le formalisme hamiltonien ordinaire ne peut donc pas être appliqué de manière immédiate aux systèmes avec contraintes. L'idée de Dirac consistait en l'introduction de nouvelles variables dans l'espace des phases de manière à améliorer les propriétés de surjectivité de la transformation de Legendre.

---

[9] À proprement parler, ce sous-ensemble n'est une sous-variété que si certaines conditions de régularité sur les contraintes sont requises. Nous ferons l'hypothèse que tel est le cas. Le terme « surface des contraintes » est couramment employé par analogie avec le cas où la sous-variété est de dimension 2.

**Hamiltonien et contraintes secondaires.** Dans le formalisme hamiltonien classique, on introduit le hamiltonien canonique, donné dans un système de coordonnées locales par l'expression

$$H_c = \sum_{a=1}^{d} v^a p_a - L(q,v).$$

Ainsi défini, le hamiltonien canonique est une fonction des positions, des vitesses et des moments. Mais on montre en utilisant la définition des moments que ce hamiltonien est, en fait, fonction seulement des positions et des moments, c'est donc une fonction sur l'espace des phases.

Cependant, dans le cas des systèmes avec contraintes, le hamiltonien n'est pas uniquement déterminé en tant que fonction des $q$ et des $p$. Afin de résoudre ce problème, Dirac introduisait un nouvel hamiltonien, $H$, appelé hamiltonien total, par addition de combinaisons linéaires des contraintes au hamiltonien canonique,

$$H = H_c + \sum_{i=1}^{I} u^i \phi_i,$$

où les $u^i = u^i(q,p)$ sont des multiplicateurs de Lagrange, c'est-à-dire des fonctions sur l'espace des phases dont la dynamique n'est pas spécifiée et qui sont totalement arbitraires. Ces multiplicateurs ne correspondent donc pas à des degrés de liberté du système. On peut montrer que, sous certaines hypothèses sur les contraintes, les équations de Hamilton associées à $H$ permettent de déterminer uniquement les vitesses en fonction des moments et des multiplicateurs. Par conséquent, ce nouvel hamiltonien est, lui, déterminé de manière unique. En résumé, il est possible de regagner l'inversibilité de la transformation de Legendre au prix de l'introduction de nouvelles variables.

On appelle variable dynamique du système une fonction sur l'espace des phases. L'évolution temporelle d'une variable dynamique est décrite à l'aide du hamiltonien total par l'equation,

$$\dot{f} = \{f, H\}.$$

À cause des contraintes primaires, le système évolue nécessairement sur la surface des contraintes, $\Sigma$. Ceci est cohérent avec la dynamique si et seulement si cette variété elle-même est laissée invariante au cours du temps, autrement dit si et seulement si les contraintes sont conservées dans le temps, $\dot{\phi}_i = 0$, c'est-à-dire, $\{\phi_i, H\} = 0$, pour $i = 1, \ldots, I$. Trois situations peuvent apparaître. Soit ces équations de cohérence sont identiquement satisfaites, soit elles entraînent une restriction sur les multiplicateurs $u^i$, soit on obtient des relations indépendantes des multiplicateurs. Dans le dernier cas, si les relations obtenues sont indépendantes des contraintes primaires, elles sont appelées des contraintes secondaires. Si on en obtient, il s'agit ensuite d'imposer leur conservation dans le temps. Ceci peut imposer des restrictions sur les multiplicateurs ou engendrer de nouvelles contraintes secondaires. Il faut alors s'assurer de leur conservation dans le temps. L'itération s'arrête lorsque les conditions de cohérence n'engendrent plus de contraintes. On a alors obtenu toutes les contraintes secondaires. La distinction entre contraintes primaires et secondaires ne comporte pas d'interprétation physique. On regroupe donc toutes les contraintes de la théorie, primaires et secondaires, en un même symbole, $\phi_j(q,p)$, avec $j = 1, \ldots, J$, où $J = I + K$, $K$ étant le nombre de contraintes secondaires.

Lorsque les $\phi_j(q,p)$, où $j = 1, \ldots, J$, définissent un ensemble complet de contraintes, les conditions de cohérence associées donnent $J$ relations, dont certaines peuvent être triviales, sur les $I$ multiplicateurs $u^i$. En interprétant ces relations en tant qu'équations linéaires non homogènes, il est possible d'écrire une solution générale pour les multiplicateurs $u^i$. Cette solution contient deux termes. Le premier, la solution particulière, est complètement fixé par les équations de cohérence, mais le second est laissé totalement arbitraire. Ainsi, même lorsque toutes les conditions de compatibilité sont satifaites, il reste un certain nombre, $P$, de fonctions arbitraires du temps dans le formalisme. Etant donné que ces fonctions arbitraires entrent dans la définition du hamiltonien total du système, l'évolution des variables dynamiques n'est pas déterminée de manière unique par leurs conditions initiales. En d'autres termes, tous les points de l'espace des phases ne correspondent pas à un état physique du système. Afin de déterminer les implications de ces remarques, il est nécessaire d'introduire la notion de contrainte de première classe.

**Contraintes de première classe et transformations de jauge.** Nous avons vu que la différence entre contraintes primaires et secondaires n'a pas de conséquences pour la dynamique du système. Par contre, une distinction fondamentale existe entre les contraintes dites de première classe et celles dites de seconde classe, et c'est à l'aide des crochets de Poisson que cette distinction peut être établie.

L'une des contraintes, définie par une fonction $\phi_j$, où $j$ est l'un des nombres $1, \ldots, J$, est dite de première classe si elle commute faiblement avec toutes les contraintes du système, c'est-à-dire si le crochet de Poisson de $\phi_j$ avec chaque contrainte du système, en restriction à la surface des contraintes, s'annule. Une propriété importante découlant de cette définition est que l'ensemble des contraintes de première classe d'un système forme une sous-algèbre de Poisson de l'algèbre de Poisson des fonctions sur l'espace des phases $\mathcal{P}$. En notant $G_n$, $n = 1, \ldots, N$, les contraintes de première classe du système, on obtient des relations, pour $n, m = 1, \ldots, N$, de la

forme

$$\{G_n, G_m\} = \sum_{p=1}^{N} f_{nm}^p G_p,$$

où, en général, les $f_{nm}^{\ p}$ sont des fonctions et non des constantes de structure.

Le nombre $N$ de contraintes de première classe d'un système est égal au nombre $P$ de fonctions arbitraires du temps apparaissant dans le hamiltonien qui ne sont pas fixées par les conditions de cohérence. Cette analogie numérique suggère que les contraintes de première classe sont reliées aux propriétés d'indétermination du système. En calculant l'évolution d'une variable dynamique à partir d'une condition initiale, on voit que deux évolutions possibles correspondant à deux choix différents de hamiltonien, c'est-à-dire à deux choix de fonctions arbitraires du temps, sont reliées par une transformation d'un certain type. En effet, si $G_n$, $n = 1, ..., N$, est une contrainte de première classe, et $f$ est une variable dynamique du système, la transformation infinitésimale[10] de $f$,

$$\delta_n f := \{f, G_n\},$$

n'affecte pas l'état physique du système, le flot correspondant relie deux valeurs de la variable dynamique $f$ qui correspondent au même état physique. C'est une transformation infinitésimale de jauge.

**Contraintes de seconde classe.** Les contraintes qui ne sont pas de première classe sont dites de seconde classe. De telles contraintes ne correspondent pas à des transformations ayant une interprétation physique. Elles sont éliminées du formalisme en introduisant une nouvelle structure d'algèbre de Poisson déterminée par un nouveau crochet, le crochet de Dirac. L'introduction de cette nouvelle structure permet d'éliminer les contraintes de seconde classe du problème et le système est alors décrit complètement par cette nouvelle structure de Poisson et les contraintes de première classe.

En résumé, les crochets de Poisson jouent un rôle prépondérant dans le traitement hamiltonien des théories de jauge. Ils permettent non seulement de déterminer la dynamique de ce type de systèmes, mais aussi, d'une part, de déterminer à travers les conditions de cohérence toutes les contraintes de la théorie et, d'autre part, de distinguer entre les contraintes de première et de seconde classe. Enfin, ils permettent de déterminer l'action des contraintes de première classe sur l'espace des phases et donc d'expliciter les transformations de jauge du système.

---

[10]Pour toutes fonctions $f$ et $g$ sur l'epace des phases, le crochet de Poisson $\{f, g\}$ s'écrit $\{f, g\} = \frac{d}{dt}|_{t=0} f \circ g_t$, où $g_t$ est le flot du champ de vecteurs dit hamiltonien associé à la fonction $g$. On dit que la fonction $g$ agit sur la fonction $f$, et le crochet de Poisson correspond donc à une action infinitésimale. Les transformations $\delta_n$ correspondent au flot sur l'espace des phases engendré par les contraintes de première classe du système.

# Quantification

Les théories quantiques sont des descriptions de la nature nées de l'étude des propriétés microscopiques des systèmes physiques. Ce ne sont cependant pas simplement des théories de l'infiniment petit, ce sont les descriptions les plus fondamentales des phénomènes physiques qui aient été développées à ce jour. Le passage de la description classique d'un système à son traitement quantique est appelé quantification (ou quantisation). Cette procédure, nécessaire pour comprendre les implications fondamentales de toute théorie classique, n'est malheureusement pas algorithmique, et il est parfois nécessaire pour quantifier un système d'utiliser des méthodes heuristiques et des approximations. De nombreux efforts de recherche en physique mathématique contemporaine sont dédiés à la question de la quantification, mais il existe à ce jour des systèmes que l'on ne sait pas quantifier, en particulier, la gravitation.

## La procédure de quantification

Il existe diverses procédures de quantification. Nous discutons une méthode particulière, la quantification canonique, approche visant à préserver autant que possible la structure formelle de la théorie classique sous-jacente. Cette méthode fut inventée par Dirac qui perçut l'analogie existant entre les crochets de Poisson et les règles de commutation dans la formulation matricielle de la mécanique quantique de Heisenberg. Il développa cette construction dans sa thèse de doctorat, publiée en 1926. Nous nous restreignons ici au cas d'un système avec un nombre fini de degrés de liberté avec lagrangien régulier, c'est-à-dire sans contraintes.

De manière générale, dans le passage de la mécanique classique à la mécanique quantique, l'espace des phases $\mathcal{P} = T^*(\mathcal{C})$ d'un système d'espace de configuration $\mathcal{C}$ est remplacé par l'espace projectif d'un espace de Hilbert complexe séparable[11], $\mathcal{H}$, appelé espace des états du système quantique. L'algèbre de Poisson des fonctions sur $\mathcal{P}$ est alors remplacée par l'algèbre des opérateurs sur l'espace de Hilbert $\mathcal{H}$. En d'autre termes, les variables dynamiques du système quantique sont des opérateurs linéaires, et les valeurs possibles d'une variable dynamique sont celles apparaissant dans le spectre[12] de l'opérateur correspondant.

Comment passer de la description classique à la description quantique du système ? Idéalement, on voudrait construire une application qui associe à toute fonction $f$ sur l'espace de phases $\mathcal{P}$ un opérateur $\iota(f)$ sur l'espace de Hilbert $\mathcal{H}$. Alors que le crochet de Poisson détermine les propriétés fondamentales du système classique, c'est le commutateur des opérateurs[13] qui détermine les propriétés fondamentales du système quantique, telles que l'évolution temporelle et les symétries. L'application $\iota$ devrait donc faire correspondre au crochet de Poisson de deux fonctions le commutateur des opérateurs correspondants, *i.e.*, être un homomorphisme d'algèbres de Lie,

$$\iota(\{f_1, f_2\}) = \frac{i}{\hbar}[\iota(f_1), \iota(f_2)], \tag{1}$$

où $i = \sqrt{-1}$ et $\hbar$ est la constante de Planck divisée par $2\pi$. On voit ici l'importance du rôle joué par le crochet de Poisson dans la quantification.

---

[11]Un espace de Hilbert est un espace vectoriel muni d'un produit scalaire hermitien. Il est dit séparable s'il possède une base dénombrable. L'espace projectif correspondant est l'ensemble des droites de cet espace passant par l'origine.

[12]Le spectre d'un opérateur est l'ensemble de ses valeurs propres.

Malheureusement une telle application n'existe pas, même dans les cas les plus simples. En effet, alors que l'algèbre des fonctions sur l'espace des phases est commutative, l'algèbre des opérateurs sur l'espace de Hilbert ne l'est pas. Il faut donc faire des choix dans l'ordre des facteurs d'un produit d'opérateurs, alors qu'un produit de variables dynamiques classiques ne dépend pas de l'ordre des facteurs. D'où une ambiguïté dans la définition du produit des opérateurs ce qui amène à affaiblir la condition imposée sur l'application de quantification en la remplaçant par une égalité au premier ordre en $\hbar$ seulement.

Il est cependant possible de construire une application satisfaisant la condition (1) de manière stricte en restreignant le domaine de l'application $\iota$ à un certain sous-espace de variables dynamiques classiques, appelées observables élémentaires. Un exemple est la représentation de Schrödinger, obtenue en choisissant pour espace de Hilbert $\mathcal{H}$ un espace de fonctions sur l'espace de configuration, décrit par les variables $q^a$, $a = 1, \ldots, d$, et pour observables élémentaires les fonctions coordonnées sur l'espace des phases, $q^a$ et $p_a$, $a = 1, \ldots, d$. Les opérateurs correspondants, $Q^a = \iota(q^a)$ et $P_a = \iota(p_a)$ agissant sur les fonctions $\psi$ appartenant à $\mathcal{H}$ sont définis par

$$(Q^a \psi)(q) = q^a \psi(q), \quad (P_a \psi)(q) = -i\hbar \frac{\partial}{\partial q^a} \psi(q).$$

On montre alors que l'application $\iota$ satisfait la propriété (1) si l'on requiert qu'elle envoie la fonction constante 1 sur l'opérateur identité sur $\mathcal{H}$. On a donc résolu au moins partiellement le problème de quantification, ce qui permet d'obtenir des informations sur les propriétés quantiques du système étudié. La question est maintenant de comprendre comment étendre ce formalisme aux systèmes avec contraintes. Il existe deux types de méthodes, la réduction de l'espace des phases et la méthode de Dirac, chacune présentant des avantages et des incovénients. Nous ferons ici l'hypothèse que le système étudié ne présente pas de contraintes de seconde classe.

---

[13]Le commutaeur de deux opérateurs $A$ et $B$ est $[A, B] = AB - BA$.

## Méthode de réduction de l'espace des phases

La première méthode consiste en l'élimination de ceux des degrés de liberté qui sont superflus en raison de l'invariance de jauge, au niveau classique, c'est-à-dire avant la procédure de quantification. Pour ce faire, on voudrait restreindre le système considéré à la surface des contraintes $\Sigma$ et considérer le système dynamique défini sur cette sous-variété de l'espace des phases total $\mathcal{P}$. Mais cette approche présente un problème, elle ne conduit pas à l'étude d'un système sur une variété symplectique de dimension inférieure. En effet la forme induite sur $\Sigma$ par la forme symplectique $\omega$ de $\mathcal{P}$ est fermée mais elle n'est pas en général non dégénérée. Par conséquent, elle n'est pas une forme symplectique et le système réduit n'est pas un système dynamique au sens usuel.

La raison de la dégénérescence de la forme induite est l'existence de transformations de jauge. Considérons les champs de vecteurs hamiltoniens associés aux contraintes de première classe $G_n$, c'est-à-dire les champs de vecteurs $X_n$ sur $\mathcal{P}$ satisfaisant

$$X_n(f) = \{f, G_n\},$$

pour toute fonction $f$ sur $\mathcal{P}$. Ces vecteurs sont tangents à la surface des contraintes $\Sigma$, et leurs restrictions à $\Sigma$ engendrent le noyau[14] de la forme induite par $\omega$ sur $\Sigma$. D'où la relation entre dégénérescence de la forme induite et transformations de jauge. Puisque la forme $\omega$ est fermée, les champs de vecteurs $X_n$ engendrent une distribution intégrable, ce qui implique que celle-ci définit un feuilletage sur $\Sigma$, et les feuilles de ce feuilletage coïncident avec les orbites des transformations de jauge.

Pour résoudre le problème soulevé initialement et afin d'obtenir une forme symplectique bien définie, il suffit de considérer le quotient $\mathcal{P}_{\text{phys}}$ de la surface des contraintes $\Sigma$ par le feuilletage ci-dessus, c'est-à-dire par les orbites des transformations de jauge. Sous certaines conditions de régularité, l'espace $\mathcal{P}_{\text{phys}}$ est une variété admettant une structure symplectique $\Omega_{\text{phys}}$ uniquement déterminée, appelé espace des phases réduit ou espace des phases physique du système contraint.

Les observables classiques étant par construction des fonctions constantes sur les orbites des transformations de jauge, on peut alors appliquer la procédure de quantification standard décrite dans la section précédente à l'espace des phases réduit $\mathcal{P}_{\text{phys}}$. Cependant, il se peut que cette méthode soit difficile à appliquer dans la pratique. D'une part, l'algèbre des observables peut être extrêmement compliquée et par conséquent difficile à quantifier. D'autre part des propriété physiques essentielles du système, telles que la localité ou l'invariance de Lorentz manifeste, peuvent être perdues par passage au système réduit. Il est donc intéressant de considérer une autre méthode, applicable lorsque la procédure de réduction ne permet pas la quantification.

---

[14] Le noyau de la forme induite par $\omega$ sur $\Sigma$ est l'ensemble des vecteurs tangents $X$ à $\Sigma$ tels que $\omega(X,Y) = 0$ pour tout vecteur $Y$ tangent à $\Sigma$.

**Procédure de quantification de Dirac**

Dans la méthode de Dirac, on ne s'emploie pas à s'affranchir des degrés de liberté de jauge au niveau de l'espace des phases, mais on quantifie d'abord toutes les variables du système en suivant la procédure décrite dans le cas des systèmes non contraints. On impose ensuite les contraintes au niveau quantique en éliminant les degrés de liberté associés à la symétrie de jauge. On obtient ainsi l'espace des états physiques du système, noté $\mathcal{H}_{\text{phys}}$. La réduction se fait ici après la quantification.

Soient $G_n$, $n = 1, ..., N$, les contraintes de première classe du système. Si les contraintes s'expriment en termes des observables élémentaires de la théorie, il est possible d'associer à chaque contrainte un opérateur $\iota(G_n)$ sur l'espace des états du système quantique $\mathcal{H}$. Les états physiques du système sont définis par leurs propriétés d'invariance par transformations de jauge, c'est-à-dire, par les transformations engendrées par les contraintes. En d'autres termes, un état est un état physique s'il appartient au noyau de chaque opérateur de contrainte $\iota(G_n)$, pour $n = 1, \ldots, N$. Notons qu'en général $\mathcal{H}_{\text{phys}}$ n'est pas un sous-espace de $\mathcal{H}$. Les états physiques sont en toute généralité des vecteurs propres généralisés des opérateurs de contrainte, et ils appartiennent au dual d'un sous-espace dense de l'espace de Hilbert de tous les états, $\mathcal{H}$.

La condition d'invariance par les transformations de jauge mène au problème de cohérence suivant. Les contraintes $G_n$, $n = 1, \ldots, N$, étant de première classe, elles engendrent par combinaisons linéaires une sous-algèbre de l'algèbre de Poisson des fonctions sur l'espace de phases, $\mathcal{P}$. À cause des ambiguïtés dues à la non-commutativité du produit des opérateurs, il est possible que, contrairement au cas classique, les opérateurs $\iota(G_n)$ correspondant aux contraintes n'engendrent pas une algèbre de Lie au sens strict mais seulement au premier ordre en $\hbar$. Les opérateurs $\iota(G_n)$ vérifient alors des relations de la forme,

$$[\iota(G_n), \iota(G_m)] = i\hbar \sum_{p=1}^{N} \iota(f_{nm}^p)\iota(G_p) + \hbar^2 D_{nm}.$$

Ceci peut avoir des conséquences importantes puisque les opérateurs $\iota(G_n)$ cessent d'être de première classe et ne peuvent donc plus être considérés comme engendrant les transformations de jauge du système quantique. On dit alors que l'invariance de jauge est brisée au niveau quantique et les opérateurs $D_{nm}$ sont appelés des anomalies de jauge. Si l'invariance de jauge est brisée par des effets d'origine quantique, il devient incohérent de sélectionner des états invariants de jauge. On voit ainsi que la procédure de Dirac ne peut être directement appliquée lorsque des anomalies sont présentes. Un exemple d'une telle situation est obtenu en considérant une corde bosonique se propageant sur un espace-temps plat. Dans ce cas, il existe une anomalie, dite anomalie de Weyl, qui s'annule si et seulement si la dimension de l'espace-temps est égale à 26.

## Le rôle des crochets de Poisson

L'exposé des méthodes d'étude des systèmes avec contraintes et de leur quantification développé dans ce chapitre montre que, malgré une histoire déjà longue, les crochets de Poisson demeurent un instrument indispensable dans les recherches sur les systèmes physiques tant classiques que quantiques. Les crochets de Poisson interviennent dans de nombreux contextes qui sont d'une importance capitale en physique théorique moderne tels que la quantification des théories de jauge, la théorie des cordes et la quantification de la relativité générale. Malgré le développement de techniques de quantification sophistiquées telles que les méthodes BRST, la quantification géométrique ou la quantification algébrique raffinée, il demeure que, pour une classe importante de cas, la quantification ne peut être effectuée qu'à travers une étude fine de la structure de Poisson de la théorie classique. Les travaux de Siméon-Denis Poisson n'ont donc pas seulement eu une importance historique en physique mathématique, ils sont à la base de développements fondamentaux en recherche contemporaine.